\documentstyle[multicol,prl,aps,epsf]{revtex}

\begin{document}

\title{Enhancement of GMR due to spin-mixing in magnetic
multilayers with a superconducting contact}
\author{F. Taddei\thanks{e-mail: f.taddei@lancaster.ac.uk},}
\address{School of Physics and Chemistry, Lancaster University,
Lancaster, LA1 4YB, UK}
\author{S. Sanvito\thanks{e-mail: ssanvito@mrl.ucsb.edu},}
\address{Materials Department, University of California, Santa Barbara,
CA 93106, USA}
\author{C.J. Lambert\thanks{e-mail: c.lambert@lancaster.ac.uk}}
\address{School of Physics and Chemistry, Lancaster University,
Lancaster, LA1 4YB, UK}
\date{\today}
\maketitle

\begin{abstract}
We study the Giant Magnetoresistance (GMR) ratio in magnetic
multiayers with
a single superconducting contact in the presence of spin-mixing processes.
It has been recently shown \cite{Fab} that the GMR ratio of magnetic multilayers
is strongly suppressed by the presence of a superconducting contact
when spin-flipping is not allowed.
In this Letter we demonstrate that the GMR ratio can be
dramatically enhanced by spin-orbit interaction and/or non-collinear magnetic
moments.
The system is described using a tight-binding model with either $s$-$p$-$d$ or
$s$-$d$
atomic orbitals per site.
\end{abstract}

\begin{multicols}{2}


Hybrid nanostructures form a fascinating melting pot for studying the interplay
between fundamental quantum phenomena, often revealing new and unexpected
physics.
One recently-recognized class of such structures, involving the coexistence of
superconducting contacts and ferromagnetic domains, has led to the identification
of a number of fundamental issues \cite{Fierz,Soul,Upd1,Pann,Petra,Upd2,Bour},
several of which are currently unresolved.
In this Letter, we examine one such issue, posed by experiments on
giant magnetoresistance (GMR) in magnetic (M) multilayers with
superconducting (S) contacts and current-perpendicular-to-the-plane (CPP).
Recognizing that the sub-gap conductance of such
structures is mediated by Andreev scattering, it was recently noted \cite{Fab}
that in the
absence of spin-flip processes, the conductance of a metallic ({\it i.e.}
diffusive) multilayer in the presence of aligned
magnetic moments is almost identical to that of the multilayer when adjacent
moments are
anti-aligned and therefore the conventional GMR ratio should be strongly
suppressed.
Since large GMR ratios are observed experimentally \cite{Pratt}, 
it is clear that even a
qualitative understanding of transport in such structures must incorporate the
effects of spin-mixing.
The aim of this Letter is to present the first theoretical description of
CPP GMR in M-multilayers with S-contacts, which incorporate spin-flip
scattering.
As sources of spin-mixing we consider both spin-orbit (SO) coupling
and non-collinear
magnetizations in adjacent magnetic layers.

The system under consideration is a disordered magnetic multilayer consisting of
an alternating sequence of magnetic layers each of length $l_M$ and non-magnetic
layers (N) of length $l_{N}$. The building block of the magnetic structure is
the bilayer [M/N] of length $l_B=l_{N}+l_M$.
The magnetic moments of even-numbered M-layers make an angle $\theta$ relative to
those of odd-numbered M-layers. Experimentally, $\theta$ can be varied by
applying an external magnetic field with
antiparallel (AP) alignment ($\theta=\pi$) typically occurring at zero field
and parallel (P) alignment ($\theta=0$)
at large enough fields.
The current flows perpendicular to the planes of the multilayer, which
makes contact with
a metallic normal lead on the left-hand side
of the multilayer and
a superconducting lead on the right-hand side.
GMR is the drastic increase in electrical conductance $G(\theta)$ that occurs when
the system switches from the AP to the P alignment with the conventional GMR
ratio defined by:
$\rho=\frac{G(0)-G(\pi)}{G(\pi)}$.

Following \cite{Fab}, the multilayer and leads are modelled using a
tight-binding Hamiltonian on a cubic lattice with hoppings to nearest neighbours.
Lattice imperfections and impurities are simulated by
adding to the on-site energies a random number in the
range $\left[ -\frac{W}{2}, +\frac{W}{2} \right]$.
The on-site Hamiltonian has the following structure:

\begin{equation}
H = \left(
\begin{array}{cccc}
H^{p\uparrow} & \mu_{xy} & \Delta & 0 \\
\mu_{xy}^\ast & H^{p\downarrow} & 0 & -\Delta \\
\Delta^\dagger & 0 & H^{h\downarrow} & -\mu_{xy}^\ast \\
0 & -\Delta^\dagger & -\mu_{xy} & H^{h\uparrow}\ \
\end{array}
\right)
\label{hamilt}
\end{equation}

\noindent
where $H^{p\uparrow (\downarrow)}$ is the Hamiltonian for up (down)-spin
particles ($s$ and $d$ bands),
$H^{h\uparrow (\downarrow)}= -H^{p\uparrow (\downarrow)\ast}$ is the
Hamiltonian for up(down)-spin holes and $\Delta$ is the superconducting
order parameter. Here $\mu_{xy}= -\mu_x + i\mu_y$, where $\mu_{x(y)}$ is
the $x(y)$-component of the exchange field $\vec{\mu}$. Note that $\vec{\mu}$ is
non-zero only for electrons in the $d$-band
in the M-layers and $\Delta$ is non-zero only in the right-hand-side
superconducting lead.
Within the tight-binding formulation, SO
interaction can be included by adding to the Hamiltonian the following term:

\begin{equation}
V_{SO} = V_1 \sum_{i,j,\alpha,s} \vec{\bf \sigma} \cdot \vec{R}_{i,j}
c_{\alpha,i}^{\sigma\dagger} c_{\alpha,j}^{-\sigma}
\label{SO}
\end{equation}

\noindent
where $V_1$ is a constant which determines the interaction strength,
$\vec{\bf \sigma}$ is a vector of Pauli matrices, $\vec{R}_{i,j}$ is the unit vector
which connects site $i$ with the neighbouring site $j$. $c_{\alpha,i}^{\sigma}$ is
the annihilation operator for electrons of spin $\sigma$ in the $\alpha$ ($s$,
$d$)-band
on site $i$. In the presence of disorder (\ref{SO}) produces spin-flip scattering
since it couples electrons with different spin on neighbouring sites.

In the presence of disorder,
to study the largest
possible sample cross sections, we consider 2 orbitals per site, which is the
minimal model
capable of reproducing scattering potential at the N/M interface and interband
scattering \cite{Ste3}.
The tight-binding parameters
are chosen to reproduce the GMR ratio and conductances obtained from an ab initio material
specific calculation for Cu/Co multilayers \cite{Ste2}.

In the presence of spin-flip scattering the
two-spin-fluid approximation does not hold and
the zero-temperature, zero-bias, normal-state Landauer formula takes
the form:

\begin{equation}
G^{NN}= \frac{e^2}{h} \sum_{\sigma \sigma^\prime} \mbox{Tr} \left\{ t^{\sigma
\sigma^\prime\dagger} t^{\sigma \sigma^\prime} \right\}
\end{equation}

\noindent
where
$t^{\sigma\sigma^\prime}$ is the matrix of transmission amplitudes for injected
$\sigma^\prime$-spin electrons in the left-hand lead into $\sigma$-spin
electrons in the right-hand lead.
When the right-hand lead is in the superconducting state, the conductance is
given by \cite{Roberto}

\begin{equation}
G^{NS}= \frac{e^2}{h}  2 \sum_{\sigma \sigma^\prime} \mbox{Tr} \left\{r_a^{\sigma
\sigma^\prime\dagger} r_a^{\sigma \sigma^\prime} \right\}
\label{AR}
\end{equation}

\noindent
where $r_a^{\sigma\sigma^\prime}$
is the Andreev reflection matrix for injected
$\sigma^\prime$-spin electrons in the left-hand lead to be reflected into
$\sigma$-spin holes.
In what follows,
the scattering amplitudes are calculated exactly by solving the Bogoliubov-de Gennes
equation using an efficient recursive Green's function technique
\cite{Fab,Ste2,Ste_tesi}.

\begin{figure}
\narrowtext
\epsfysize=6cm
\epsfxsize=8cm
\epsffile{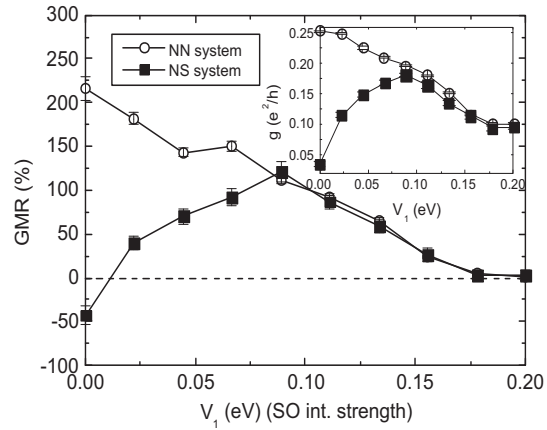}
\caption{ \label{GMR_SO}
Conventional GMR ratio as a function of the SO interaction strength for NN and NS cases.
Results correspond to disordered multilayers ($W=0.6$ eV) comprising
40 bilayers with $l_M=15$ and $l_N=8$. Samples are formed by repeating
(3$\times$3) disordered unit cells
in the transverse plane and summing over 25 k-points in the 2-dimensional Brillouin
zone. The points are an average over 20 disorder realizations and
the error bars represent the standard deviations from the mean.
In the insert, comparison between the conductances in the P alignment for the NN
and NS cases as functions of the SO interaction strength.}
\end{figure}

\begin{figure}
\narrowtext
\epsfysize=6cm
\epsfxsize=8cm
\epsffile{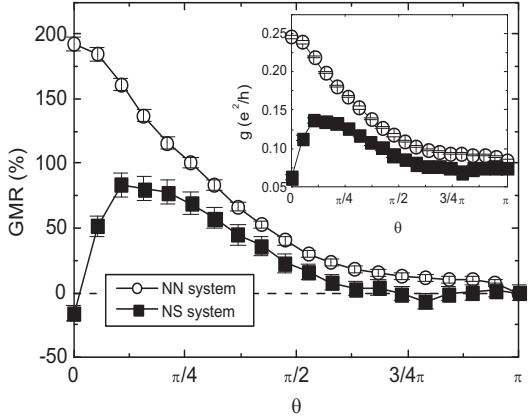}
\caption{ \label{dis_GMR_40}
$\theta$-dependent GMR ratio for NN and NS cases in the absence of SO coupling.
Results correspond to disordered multilayers ($W=0.6$ eV)
of 22 bilayers with $l_M=30$, $l_N=16$, considering a (3$\times$3) unit cell
in the transverse plane, and a sum over 25 k-points in the 2-dimensional Brillouin
zone. The points are the average over 50 realizations of disorder.
In the insert, $\theta$-dependent conductance for NN and NS cases in the absence
of SO coupling.}
\end{figure}

We shall now turn to
the central results of this
Letter, namely that in the presence of
strong-enough spin-mixing, either produced by SO
coupling or non-collinear moments, GMR in the presence of a S-contact
approaches that of
two normal contacts, with values of $\rho$ of the order of 100~\%.
First consider the effect of SO coupling within the multilayer. Fig.
\ref{GMR_SO} shows the conventional GMR ratio as a function of the SO
interaction strength $V_1$ for a disordered multilayer of 40 bilayers, with
$l_M=15$ and $l_N=8$. As expected, in the NN case $\rho$ decreases
monotonically from $\simeq 200$~\% at $V_1=0$, to zero at large $V_1$ ($\simeq 0.17$
eV). In contrast for
the NS case, $\rho$ initially increases with increasing $V_1$, eventually joining the
NN curve at $V_1\simeq 0.08$ eV.
As a second source of spin-mixing,
consider the effect of non-collinear magnetic moments when $V_1=0$. Fig.
\ref{dis_GMR_40} shows the $\theta$-dependence of the GMR ratio defined as
$\rho(\theta) = \frac{ G(\theta) - G(\pi)}{G(\pi)}$. Whereas in the NN case
$\rho(\theta)$
decreases monotonically with increasing $\theta$, in the NS case $\rho(\theta)$
exhibits a pronounced
maximum around $\theta=\pi/8$.

To understand these results, first consider the case of
non-collinear moments.
In the presence of two normal-metallic contacts, the conductance $G(\theta)$ 
has been theoretically
studied in \cite{ang_th1,ang_th2} where it is predicted that
$G(\theta)-G(\pi)$ tends monotonically to
zero as $\theta$ varies from $0$ to $\pi$. In addition,
the dependence of
the resistance on the angle $\theta$ has been experimentally found \cite{ang_ex} to contain a
term proportional to cos$^2(\theta/2)$ and a second term proportional to
cos$^4(\theta/2)$.
In the presence of a S-contact,
where $G(0) \simeq G(\pi)$, this behaviour is drastically changed
by the presence of an extremum which
occurs at some intermediate angle
$\theta_c$, the value of which depends on the interplay between competing
effects. Since $\theta(H)$ is a function of the applied magnetic field $H$, the
presence of a S-contact introduces a new characteristic field $H_c$ for which
$\theta(H_c) = \theta_c$.
For a disordered multilayer of 22 bilayers, with $l_M =30$, $l_N =16$, the insert
in Fig.
\ref{dis_GMR_40} shows the $\theta$-dependence of the conductance divided by the
number of open channels in the normal lead. As expected $G^{NN} (\theta)$ is a
monotonic function of $\theta$, whereas $G^{NS} (\theta)$ possesses an extremum
at $\theta_c \simeq \pi/8$. To understand why the extremum is a maximum, recall
that for $\theta =0$ or $\theta =\pi$, when spin is conserved, current flows when
a right-going (spin $\sigma$) electron passes through the multilayer, Andreev
reflects as a left-going (spin $-\sigma$) hole, which retraverses the
multilayer. A M-layer whose moment is aligned with the spin of the incident
electron is anti-aligned with the spin of the outgoing hole and consequently the
number of aligned and anti-aligned M-layers encountered by a given
quasi-particle is the same for both $\theta =0$ and $\theta =\pi$ (only the
order differs).
When the elastic mean free path is comparable with
the total multilayer length, the
resistance of traversed layers add in series and therefore, apart from
small differences due to interference effects \cite{Ste4},
$G^{NS} (0) \simeq G^{NS} (\pi)$. 
Furthermore, since a quasi-particle must necessarily traverse
regions in which it is a minority spin,
both $G^{NS} (0)$ and $G^{NS} (\pi)$ are low-conductance
states.
In contrast, as $\theta$ increases from zero, this conductance bottleneck is
removed, because an Andreev reflected minority hole can spin-convert to a
majority hole, thereby avoiding anti-aligned moments on its return journey. Of
course this initial increase in $G^{NS} (\theta)$ is eventually overcome by the
usual GMR effect which decreases $G^{NS} (\theta)$ as $\theta \rightarrow \pi$, thereby
producing an overall maximum.

\begin{figure}
\narrowtext
\epsfysize=5cm
\epsfxsize=5cm
\epsffile{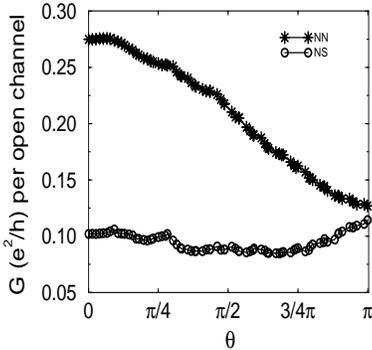}
\caption{ \label{clean_spd}
$\theta$-dependent conductance for NN and NS cases in the absence of SO coupling
for a clean multilayer. The multilayer is modelled by a material-specific
$spd$-band Hamiltonian
(see Ref. [1]),
with Co as M-material, Cu as
N-material and Pb as S-material. $l_M=7$ and $l_N=10$, considering a
(1$\times$1) unit cell in the transverse plane, summing over about 5000 k-points in
the 2-dimensional Brillouin zone.}
\end{figure}

\begin{figure}
\narrowtext
\epsfysize=5cm
\epsfxsize=5cm
\epsffile{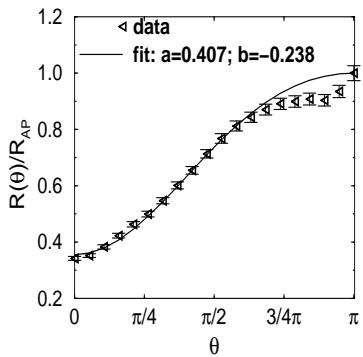}
\caption{ \label{ang_fit}
Plot of the ratio $\frac{R(\theta)}{R(\pi)}$ along with the best fit to the
function (\ref{ratio}). The value of the fitting parameters are: $a=0.407$,
$b=-0.238$. Results correspond to a disordered multilayer with the same parameters
as in Fig. \ref{dis_GMR_40}.}
\end{figure}

In the absence of disorder, the nature of the extremum is determined by
interface scattering and band structure. To illustrate this
consider a clean multilayer which is perfectly periodic and therefore
the variation of the
conductance with $\theta$
arises from tuning of the ballistic
spin-filtering by the structure.
Fig. \ref{clean_spd} shows the conductance divided by the
number of open channels in the left-hand-side normal lead.
As expected, $G^{NN}(\theta)$ is a monotonic function of
$\theta$, whereas $G^{NS}(\theta)$ exhibits a minimum around $\pi /2$ and
then increases.
(In this case, translational invariance in
the transverse direction allowed us to use a full {\it ab initio}, $spd$
Hamiltonian to obtain the results of Fig. \ref{clean_spd}.)
In the NN case, the dependence of the multilayer
resistance on $\theta$ predicted by our model is in good agreement with
experiment \cite{ang_ex}. In Ref. \cite{ang_ex} the ratio between the resistance at a given
$\theta$ and the resistance with AP alignment
has been found to fit the following function:
\begin{equation}
\frac{R(\theta)}{R(\pi)} = 1-a \cos^2 (\theta/2) + b \cos^4 (\theta/2)
\label{ratio}
\end{equation}
where $a$ and $b$ are fitting constants.
In Fig. \ref{ang_fit} we show the plot of such a ratio for the disordered multilayer
considered above, along with the best fit to function (\ref{ratio}). In addition
we also checked that this ratio cannot be fitted with the same accuracy
assuming a pure dependence on cos$^2 (\theta/2)$ ({\it i.e.} with $b=0$).
For $G^{NS}(\theta)$ however, no such analytic results currently exist.

\begin{figure}
\narrowtext
\epsfysize=7cm
\epsfxsize=8cm
\epsffile{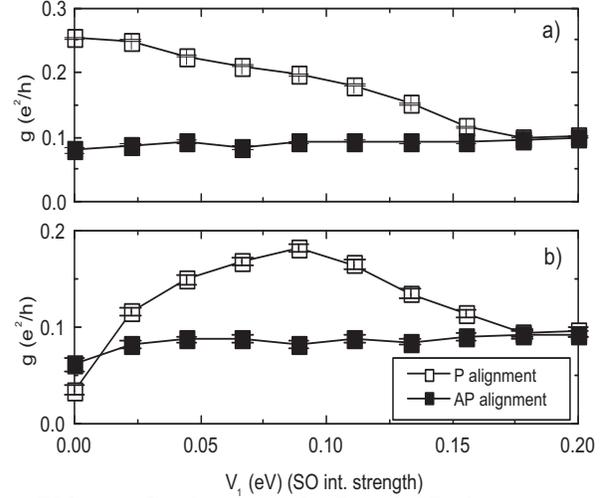}
\caption{ \label{G_SO_NN}
Conductances for P and AP alignments as functions of the SO interaction
strength for the NN case (a) and for the NS case (b).
Results correspond to a disordered multilayer with the same parameters as in Fig.
\ref{GMR_SO}.}
\end{figure}

Let us now turn attention to the effect of SO coupling.
Figs. \ref{G_SO_NN}a and \ref{G_SO_NN}b show the conductances as a function of the SO
strength $V_1$ for, respectively, the NN and the NS case.
In the NN case (Fig. \ref{G_SO_NN}a) $G_P^{NN}$ decreases as $V_1$ increases and eventually joins
the curve for
$G_{AP}^{NN}$. This can be understood in terms of the heuristic model
presented in Ref. \cite{Fab}, because, as the SO
strength increases, the
average length required for a spin to flip (spin relaxation length
$\lambda_{sf}$) gets shorter. Therefore in the P alignment, an injected majority
electron travels through the multilayer for a length $\lambda_{sf}$ before being
scattered into a minority spin, thereby producing a decrease in the
conductance. This suggests that the value of $V_1$ for which $G^{NN}_P \simeq
G^{NN}_{AP}$ corresponds to a spin relaxation length $\lambda_{sf}$ close to
the period $l_B$ of the multilayer.
We have carried out a range of simulations which show that this value of $V_1$
does not depend on the overall length of the multilayer, but decreases with
increasing $l_B$.
As expected, the conductance with AP alignment does not change
significantly with $V_1$.

In the NS case (Fig. \ref{G_SO_NN}b) the conductance in the P aligned state rapidly increases
with $V_1$, reaching a maximum and thereafter decreases,
eventually joining the
curve for the AP configuration. Clearly the enhancement in $G_P^{NS}$ is
produced by the onset of spin-flip scattering.
The 
abrupt increase is understandable, since even a
relatively small probability for spin flipping opens a highly conductive
``channel''
if the spin-flip events take place in the vicinity of the interface.
As one can see in the insert of Fig. \ref{GMR_SO} for larger values of $V_1$, the conductance
$G_P^{NS}$ joins $G_P^{NN}$ and
together they decrease thereafter.
As in the normal case $G_{AP}^{NS}$ depends weakly on $V_1$.
The value $V_1\simeq 0.08$, at which
$G^{NS}$ is maximum, corresponds
to a spin relaxation length close to the total length of the multilayer and, as
expected, separate simulations show that this value of $V_1$ decreases with
increasing total length.
Similarly the value of $V_1$ at which the GMR ratio (of Fig. \ref{GMR_SO}) vanishes
corresponds to a spin-relaxation length of the order the bilayer thickness
$l_B$ and is independent of the total length of the system.

In conclusion, we have demonstrated that spin-mixing plays a crucial r\^ole in
determining both the qualitative and quantitative features of GMR in magnetic
multilayers with a S-contact. In contrast with the normal case, where
spin-mixing suppresses GMR, we find that the GMR ratio can be dramatically
enhanced by the presence of spin-orbit interactions and/or non-collinear
magnetic moments. In experiments carried out to-date, the presence of large
spin-orbit scattering \cite{Pratt2} presumably masks the mechanism shown in Fig.
\ref{dis_GMR_40}, which is predicted to be a generic feature in the absence of
other spin-mixing processes. This suggests that lighter metals and
superconductors would be more appropriate for observing the new extrema
predicted in this Letter.
Finally we note that for the future it would be of interest to examine
spin-mixing in non-diffusive NS structures such as clean spin-valves
\cite{france}, where the GMR ratio can be non-zero or negative, even in the
absence of spin-flip processes.

\end{multicols}

\end{document}